\documentclass[12pt]{JHEP3}

\usepackage{amssymb}
\usepackage{epsfig,multicol,amsmath}
\newcommand\fverb{\setbox\pippobox=\hbox\bgroup\verb}
\newcommand\fverbdo{\egroup\medskip\noindent%
\fbox{\unhbox\pippobox}\ }			
\newcommand\fverbit{\egroup\item[\fbox{\unhbox\pippobox}]}
\newbox\pippobox
\def\d2bar{$\overline{\mbox D2}$}
\title{Super Jackstraws and Super Waterwheels}
\author{Jin-Ho Cho\\
Center for Quantum Space Time, Sogang University, Seoul, 121-742, Korea\\
              \&
Department of Physics, Kyung Hee University, Seoul, 130-701, Korea\\
E-mail: \email{cho.jinho@gmail.com}}

\preprint{\hepth{06mmxxx}} 

\abstract{We construct various new BPS states of D-branes preserving $8$ supersymmetries. These include super Jackstraws (a bunch of scattered D- or $(p,\,q)$-strings preserving supersymmetries), and super waterwheels (a number of D$2$-branes intersecting at generic angles on parallel lines while preserving supersymmetries). Super D-Jackstraws are scattered in various dimensions but are dynamical with all their intersections following a common null direction. Meanwhile, super $(p,\,q)$-Jackstraws form a planar static configuration.  We show that the $SO(2)$ subgroup of $SL(2,\,\mathbf{R})$, the group of classical S-duality transformations in IIB theory, can be used to generate this latter configuration of variously charged $(p,\,q)$-strings intersecting at various angles. The waterwheel configuration of D$2$-branes preserves $8$ supersymmetries as long as the `critical' Born-Infeld electric fields are along the common direction. }


\begin{document} 
\section{Introduction and Summary}


Static D-strings intersecting at general angles do not preserve any supersymmetry of the vacuum. The inter-strings extending over two D-strings contain tachyonic modes so that those D-strings tend to rearrange themselves reducing their energy \cite{polchinski}. (See also Refs. \cite{Cho:2003qu} and \cite{Cho:2005aj} for the argument in the T-dual setup.) Two static D-strings can preserve 16 supersymmetries only when they are parallel.

Static higher dimensional D-branes keep partial supersymmetries even when they are not parallel. For example, two D$2$-branes, intersecting at angles, $\pi-\phi_{1}$ and $\pi-\phi_{2}$, in the planes $(x^{1},\,x^{2})$- and $(x^{3},\,x^{4})$-planes respectively, preserve $8$ supersymmetries when $\phi_{1}+\phi_{2}=0$. In the case of two D$3$- or higher branes, we have more possibilities of supersymmetric configurations. They are T-dual to various BPS bound states of D$0$-D$(2k)$ with an appropriate arrangement of Born-Infeld (BI) fields over the world-volumes of D$(2k)$-branes \cite{Mihailescu:2000dn}\cite{Witten:2000mf}.

One way of giving supersymmetries to two intersecting D-strings is to make them dynamical so that the intersection point move at the light speed \cite{Bachas:2002qt}\cite{Myers:2002bk}\cite{Cho:2002ga}. (The resulting configuration is thus called `null scissors'.) Another non-trivial example of dynamical but supersymmetric configuration of D-strings is the D-helix configuration (a coil made of a D-string) \cite{Cho:2001ys}. D-helix keeps $8$ supersymmetries when it is in motion on its axis at the light speed. Without this null motion, the nontrivial helical profile cannot sustain the tension of the D-string \cite{Lunin:2001fv}. D-helix is T-dual to a supertube, a tubular configuration of a static D$2$-brane with appropriate BI fields over its world-volume \cite{Mateos:2001qs}.

In this paper, we find various new configurations of D-branes (and of their T- or S-dual cousins) preserving $8$ supersymmetries. In particular, we obtain a planar configuration of various static $(p,\,q)$-strings intersecting at generic angles, nevertheless preserving $8$ supersymmetries. We call it super $(p,\,q)$-Jackstraws. The way they preserve supersymmetries despite their generic posing angles is to tune their charges (NS-NS charge, $p$, and R-R charge, $q$) according to their posing angles. On the contrary, the posing angle of a given additional $(p,\,q)$-string is determined by the very element of $SO(2)$ subgroup of $SL(2,\,\mathbf{R})$ (the group of classical S-dual transformations in type IIB theory \cite{Schwarz:1995dk}), that gives the specific charge, acting on a reference D-string.   

We start from generalizing the null scissors by arranging additional moving D-strings without breaking supersymmetry. This exactly amounts to playing the old game `Jackstraws' in reverse order. We find the supersymmetry condition on the tilting angles and the rapidities of D-strings by checking the compatibility among the supercharges preserved by each of D-strings. In the resulting dynamical, but supersymmetric configuration, D-strings are posed so that all their intersection points follow parallel null lines in arbitrary dimensions. Therefore we call the specific configuration super D-Jackstraws. Various combinations of T-duality transformations results in other interesting supersymmetric configurations of D-branes. These include super waterwheels (D$2$-branes intersecting at generic angles on the wheel axis with BI fields on their world-volumes) and super $(p,\,q)$-Jackstraws. 

To be more specific about $SO(2)$ transformation, let us look back on the role of $SL(2,\,\mathbf{R})$ on the solution of IIB supergravity.
Type IIB theory is self-dual under S-dual transformation. At low energy, the transformations are realized as $SL(2,\,\mathbf{R})$ acting on the doublet $(H^{(3)},\,F^{(3)})$ composed of NS-NS $3$-form and R-R $3$-form field strengths, and  on the complex scalar field $\tau=\chi+i e^{-\phi}$ composed of R-R 0-form, $\chi$, and NS-NS 0-form field, $\phi$, in the fashion $\grave{a}\,\,la$ M\"{o}bius transformation on a Rieman surface \cite{Schwarz:1995dk}. In general, the transformation mixes NS-NS-charge and R-R-charge to produce $(p,\,q)$-strings. 

The vacuum of IIB theory is characterized by the asymptotic value $\tau_0=\chi_0+i e^{-\phi_0}$ of the scalar fields, and $SL(2,\,\mathbf{R})$ transforms those vacuum moduli to one another. Especially under the subgroup $SO(2)$ of S-dual transformation group, the specific vacuum with $\chi_{0}=\phi_{0}=0$ is a fixed point, that is to say, the asymptotic value, $\tau_ 0=i$, is invariant. Hence, given a supergravity solution constructed on this particular vacuum, we get another by acting an element of $SO(2)$ subgroup. For example, from a supergravity solution of a stack of D-strings, say a $(0,\,q)$-string, we draw the solution of a $(p',\,q')$-string with the same value of $\tau_{0}=i$, just by acting an appropriate element of $SO(2)$;
\begin{eqnarray}\label{so(2)trans}
\left(\begin{array}{c}p' \\q'\end{array}\right)&=&\left(\begin{array}{rr}\cos{\theta} &\quad -\sin{\theta} \\\sin{\theta} & \cos{\theta}\end{array}\right)\left(\begin{array}{c}0 \\q\end{array}\right) \nonumber\\
&=&\left(\begin{array}{r}-q \sin{\theta} \\q \cos{\theta}\end{array}\right).
\end{eqnarray}  

We claim that if we pose a $(p',\,q')$-string and a $(0,\,q)$-string so that they intersect at the very angle $\theta$ in the above, then $8$ supersymmetries will be preserved. The same is true even when we add more $(p_{(a)},\,q_{(a)})$-strings ($a=1,\,2,\cdots$) as long as they compose a planar configuration making angles $\tan{\theta_{(a)}}=-p_{(a)}/q_{(a)}$ with respect to the referential $(0,\,q)$-string. 

\FIGURE{
\epsfig{file=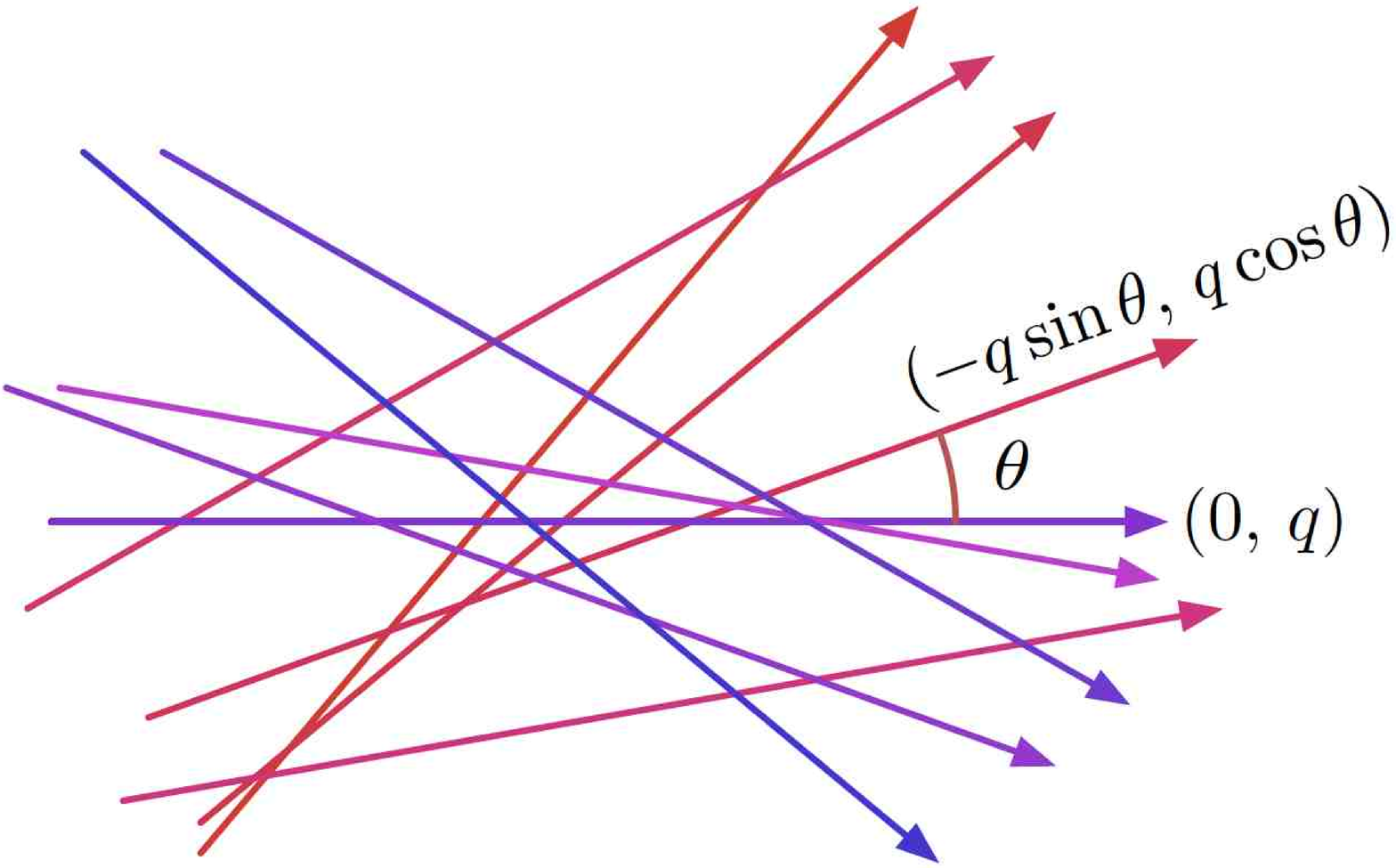,width=9cm} 
\caption{\small A static configuration of $(p,\,q)$-strings preserving $8$ supersymmetries. The tilting angle of each string is determined by its carrying charge $(p,\,q)$. Different colors denote different $(p,\,q)$ charges.}
\label{figure3}}

%


This paper is organized as follows. As a preliminary to the main topic, we review in Sec. \ref{secii} that a D-string with BI electric field on its world-sheet is nothing but BPS composite state of $p$ fundamental strings and a D-string. We postpone the details about the basic tool we adopt for the supersymmetry analysis, to Appendix \ref{appi}. There, we focus on two intersecting D-strings and explain the supersymmetry condition in our context. We obtain in Sec. \ref{seciii} the condition for two moving D-strings intersecting at a generic angle to preserve supersymmetries. We show how the supersymmetric null scissors appear in the present context. In Sec. \ref{seciv}, we extend the analysis to the cases of three D-strings in motion on a plane. In Sec. \ref{secv}, we show that two D-strings moving in two different planes that are non-intersecting cannot preserve any supersymmetry. Sec. \ref{secvi} is devoted to the case of three moving D-strings which are not necessarily in a plane. We show that they preserve $8$ supersymmetries when all the intersections follow null lines in parallel. This result is valid irrespective of the number of D-strings. In Sec. \ref{secvii}, by taking T-duality on the super D-Jackstraws, we obtain super waterwheels configuration. In order to keep supersymmetries, there should be the critical valued BI electric field (along the common direction) on each D$2$-brane. In Sec. \ref{secviii}, taking sequential T-dualities on a specific configuration of super D-Jackstraws, we arrive at a planar supersymmetric configuration of static $(p,\,q)$-strings intersecting at generic angles, that is, super $(p,\,q)$-Jackstraws. Sec. \ref{secix} concludes the paper with some discussions on the properties of the super Jackstraws.  

\section{Gauged D-strings as a $(p,\,q)$-string}\label{secii}
A D-string with the electric field on its world-volume can be considered as a $(p,\,1)$-string, i.e., the composite of $p$ fundamental strings and a single D-string \cite{Callan:1995xx}. In fact, from Born-Infeld Lagrangian of a D-string (compactified on a circle of radius $l$),
\begin{equation}\label{}
\mathcal{L}=- \frac{l}{\lambda}\sqrt{1-E^2}=- \frac{l}{\lambda}\sqrt{\frac{l^2}{l^2+\lambda^2\Pi^2}},
\end{equation}  
with
\begin{equation}\label{}
\Pi= \frac{\partial \mathcal{L}}{\partial \dot{A}}= -\frac{lE}{\lambda\sqrt{1-E^2}},
\end{equation}  
one can derive
\begin{equation}\label{pqtension}
\mathcal{H}=\Pi \dot{A}-\mathcal{L}=\sqrt{\frac{l^2}{\lambda^{2}}+\Pi^2}=l\sqrt{\frac{1}{\lambda^2}+p^2}.
\end{equation}  
Here we set the string tension $1/2\pi\alpha'=1$ and used the quantization of the momentum $\Pi=-p\,l$ with $p\in \mathbf{Z}$. The parameter $\lambda$ represents the string coupling $e^{\phi_{0}}$. The field $E$ is the component $\mathcal{F}^{01}$ of the gauge invariant two-form, $\mathcal{F}^{(2)}=dA^{(1)}-B^{(2)}$, that is, a combination of the $U(1)$ gauge field strength $dA^{(1)}$ on the D-string worldvolume and NS-NS bulk field $B^{(2)}$ pulled back onto the same world volume. When $p=0$, the tension becomes that of a D-string; $\mathcal{H}/l=1/\lambda$. Meanwhile, in the strong coupling limit of $\lambda\rightarrow\infty$, it becomes that of a fundamental string. 

The way to see that the gauged D-string carries NS-NS charge is to vary the supergravity action with respect to the field $B^{(2)}$ \cite{Witten:1995im}:
\begin{equation}\label{}
I=-\int\,d^{10} x \sqrt{-G} \frac{1}{2\lambda^2}\vert dB^{(2)}\vert^2 -\frac{1}{\lambda}\int d^2 x \sqrt{1-E^2}.
\end{equation}   
Since $\mathcal{F}^{(2)}=dA^{(1)}-B^{(2)}$, we will get a source term in the equation of motion for NS-NS three-form $H^{(3)}=dB^{(2)}$. Therefore the tension (\ref{pqtension}) of the gauged string is nothing but that of a $(p,\,1)$-string. By superposing $\bar{q}$ BPS $(p,\,1)$-strings, so that $\bar{p}=\bar{q}\,p$, we obtain a more generic $(\bar{p},\,\bar{q})$-string.

\section{Intersecting D-strings in Motion}\label{seciii}

In this section, we obtain the condition of preserving supersymmetries for two intersecting D-strings. The basic tool used for the supersymmetry analysis is explained in Appendix \ref{appi}.
Let us arrange two D-strings so that they are tilted at angle $\theta_{1,2}$ with respect to $x^{1}$-axis and are moving transversely with the rapidity $\gamma_{1,2}$ respectively. The supercharge concerning each D-string is of the form
\begin{equation}\label{}
Q+\beta^{2}_{a}\beta^{\bot}_{2}\tilde{Q},\qquad \beta^{2}_{a}=\rho(-\theta_{a})\rho(-\gamma_{a})
\beta^{2}\rho(\gamma_{a})\rho(\theta_{a}), \qquad (a=1,\,2).
\end{equation}     
Fig. \ref{figure3} shows the configuration. The spinor states invariant under both super charges are $16$ component spinors $\epsilon$ satisfying the compatibility condition
\begin{eqnarray}\label{}
(\beta^{2}_{1})^{-1}\beta^{2}_{2}\,\epsilon&=&\rho(-\theta_{1})\rho(-\gamma_{2})
(-1)\beta^{2}\rho(\gamma_{1})\rho(\theta_{1})\rho(-\theta_{2})\rho(-\gamma_{2})
\beta^{2}\rho(\gamma_{2})\rho(\theta_{2})\epsilon \nonumber\\
&=&\left(-\sinh{\gamma_{1}}\Gamma^{0}- \sin{\theta_{1}}\cosh{\gamma_{1}}\Gamma^{1}+\cos{\theta_{1}}\cosh{\gamma_{1}}\Gamma^{2}\right)\cdot \nonumber\\
&& \left(-\sinh{\gamma_{2}}\Gamma^{0}-\sin{\theta_{2}} \cosh{\gamma_{2}}\Gamma^{1}+\cos{\theta_{2}}\cosh{\gamma_{2}}\Gamma^{2}\right)\nonumber\\
&=&\left[\mathbf{1}\left(\cos{(\theta_{1}-\theta_{2})}\cosh{\gamma_{1}}\cosh{\gamma_{2}}-\sinh{\gamma_{1}}\sinh{\gamma_{2}} \right)\right. \nonumber\\
&&+\Gamma^{01}\left(\sin{\theta_{2}}\sinh{\gamma_{1}}\cosh{\gamma_{2}}-\sin{\theta_{1}}\cosh{\gamma_{1}}\sinh{\gamma_{2}} \right) \nonumber\\
&&+\Gamma^{02}\left( \cos{\theta_{1}}\cosh{\gamma_{1}}\sinh{\gamma_{2}}-\cos{\theta_{2}}\sinh{\gamma_{1}}\cosh{\gamma_{2}}\right)\nonumber\\
&&\left. - \Gamma^{12}\sin{(\theta_{1}-\theta_{2})}\cosh{\gamma_{1}}\cosh{\gamma_{2}}\right] \,\epsilon=\epsilon.
\end{eqnarray}
Since the eigenoperator and the operators $S_{0}$ and $S_{1}$ do not commute\footnote{These are defined in the Appendix.}, the eigenvector (here, the spinor $\epsilon$) must be some combination of the above basis elements. Let
\begin{eqnarray}\label{}
\epsilon&=&a|1,\,1,\,2s_{2},\,2s_{3},\,2s_{4}>+b|1,\,-1,\,2s_{2},\,2s_{3},\,2s_{4}>\nonumber\\
&&+c|-1,\,1,\,2s_{2},\,2s_{3},\,2s_{4}>+d|-1,\,-1,\,2s_{2},\,2s_{3},\,2s_{4}>\nonumber\\
&\equiv&(a,\,b,\,c,\,d),
\end{eqnarray}   
in terms of which the eigenvalue equation can be rewritten as  
\begin{eqnarray}\label{}
&&\left(\cos{(\theta_{1}-\theta_{2})}\cosh{\gamma_{1}}\cosh{\gamma_{2}}-\sinh{\gamma_{1}}\sinh{\gamma_{2}}-1 \right)(a,\,b,\,c,\,d)\nonumber\\
&&+\left(\sin{\theta_{2}}\sinh{\gamma_{1}}\cosh{\gamma_{2}}-\sin{\theta_{1}}\cosh{\gamma_{1}}\sinh{\gamma_{2}} \right)(d,\,c,\,b,\,a)\nonumber\\
 &&-i\left( \cos{\theta_{1}}\cosh{\gamma_{1}}\sinh{\gamma_{2}}-\cos{\theta_{2}}\sinh{\gamma_{1}}\cosh{\gamma_{2}}\right)(d,\,-c,\,b,\,-a)\nonumber\\
 &&-i\sin{(\theta_{1}-\theta_{2})}\cosh{\gamma_{1}}\cosh{\gamma_{2}}(a,\,-b,\,c,\,-d) =0.
\end{eqnarray}
 Nontrivial spinors exist only when the characteristic equation is satisfied;
 \begin{equation}\label{consistency}
\cos{(\theta_{1}-\theta_{2})}\cosh{\gamma_{1}}\cosh{\gamma_{2}}-\sinh{\gamma_{1}}\sinh{\gamma_{2}}=1
\end{equation} 
 Then $a$ and $d$ are related and so are $b$ and $c$, whence $8$ independent spinors satisfy the above Killing spinor equation.
  
\FIGURE{
\epsfig{file=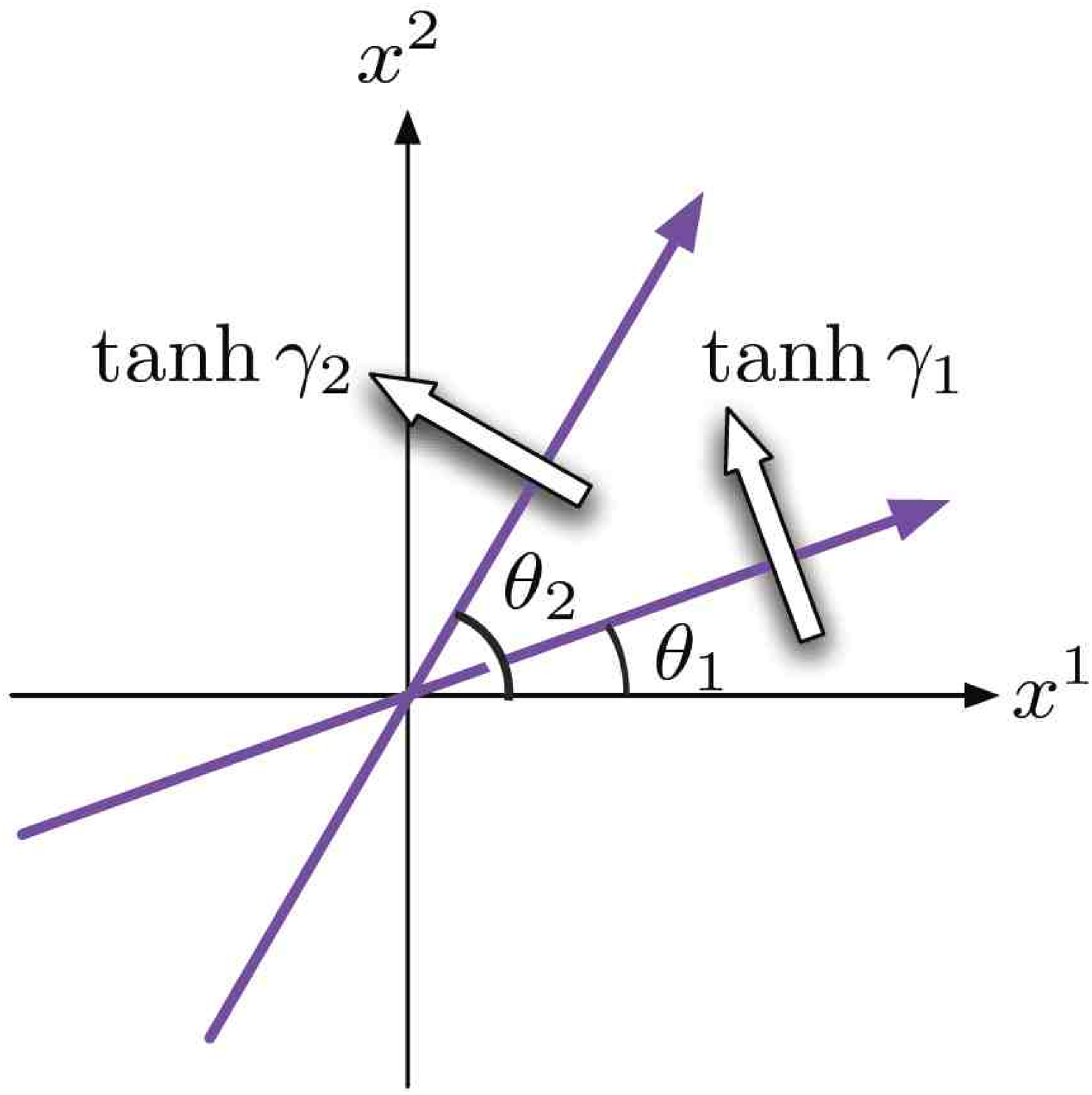,width=9cm} 
\caption{\small two intersecting D-strings (bold solid arrows) intersecting at an angle $(\theta_{2}-\theta_{1})$ and boosted with the rapidity $\gamma_{1}$ and $\gamma_{2}$}
\label{figure3} }

%


The geometric meaning of the above condition is that the intersection point of two moving D-strings moves at the light speed. This can be seen from Fig. \ref{figure4} easily, which depicts the profiles of two D-strings tilted at angle $\theta_{1}$ and $\theta_{2}$ and moving with the speed $\tanh\gamma_{1}$ and $\tanh\gamma_{2}$ respectively. During time $\delta t$, the intersection point follows the dotted line, whose distance $l$ is
\begin{eqnarray}\label{}
l&=&\left(\left(l_{2}-l_{1} \right)^{2}+\left(\delta t \tanh{\gamma_{2}} \right)^{2}   \right)^{\frac{1}{2}}  \nonumber\\
&=&\left(\frac{\tanh^{2}{\gamma_{1}}+\tanh^{2}{\gamma_{2}}-2 \cos{(\theta_{2}-\theta_{1})}\tanh{\gamma_{1}}\tanh{\gamma_{2}}}{\sin^{2}{(\theta_{2}-\theta_{1})}} \right)^{ \frac{1}{2}}\delta t \nonumber\\
&=&\delta t.
\end{eqnarray}
In the second line, $l_{1}\sin{(\theta_{2}-\theta_{1})}=\delta t\tanh{\gamma_{1}}$ and  $l_{2}\tan{(\theta_{2}-\theta_{1})}=\delta t\tanh{\gamma_{2}}$ were used, while in the last line, the above consistency condition (\ref{consistency}) was used.

\FIGURE{
\epsfig{file=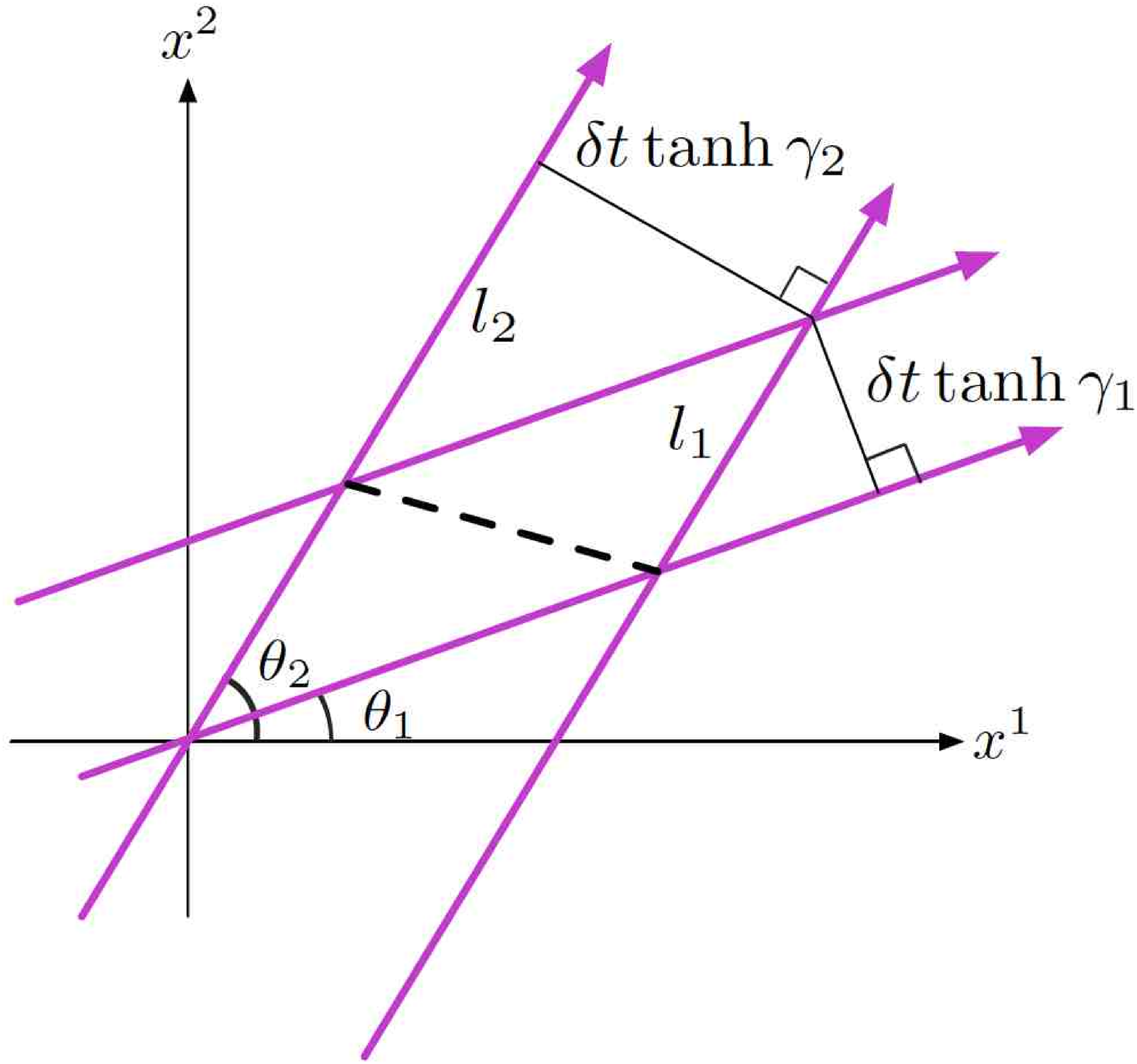,width=9cm} 
\caption{\small The profile made by two D-strings during time $\delta t$.}
\label{figure4}}

%


\section{Three D-strings}\label{seciv}
Let us extend the previous argument on the supersymmetric configuration of two D-strings to the case of three D-strings. In addition to the previous two D-strings, we arrange one more static D-string along $x^{1}$-axis (with its tilting angle and rapidity vanishing). The compatibility conditions for the supercharges concerning each D-string are 
\begin{eqnarray}\label{}
\begin{array}{ll}
\cos{(\theta_{1}-\theta_{2})}\cosh{\gamma_{1}}\cosh{\gamma_{2}}- \sinh{\gamma_{1}}\sinh{\gamma_{2}}=1 & (\mbox{between string 1 and 2}) \\ 
\cos{\theta_{1}}\cosh{\gamma_{1}}=1& (\mbox{between string 1 and 3}) \\\cos{\theta_{2}}\cosh{\gamma_{2}}=1 & (\mbox{between string 2 and 3})\end{array}
\end{eqnarray} 
We first note that the angles are restricted to the range
\begin{equation}\label{}
- \frac{\pi}{2}<\,\theta_{1},\,\theta_{2}\,< \frac{\pi}{2}
\end{equation}  
in order to make sense of the above equations. This means that the intersecting angle of any two D-strings should be less than $\pi/2$. Another thing clear from the equations is that the intersecting angle vanishes if and only if the rapidity difference vanishes.

The above three conditions are not fully independent. In fact, from the second and the third equations, one can derive $\sin{\theta_{i}}\cosh{\gamma_{i}}=\pm \sinh{\gamma_{i}}$ $(i=1,2)$. If the relation is the same (either with `$+$' or `$-$') for both D-strings, the first equation is automatically satisfied. This is when the intersecting point of the first and the third string and that of the second and the third string moves at the light speed in the same direction along the third string\footnote{Actually this is not that simple. Given two solutions $\epsilon_{1}$ and $\epsilon_{2}$ satisfying $\mathcal{O}_{i}\epsilon_{i}=\epsilon_{i}$ respectively for $i=1,\,2$, it is never trivial to find the solution $\epsilon$ satisfying $\mathcal{O}^{-1}_{1}\mathcal{O}_{2}\epsilon=\epsilon$. For the cases at hand, $\epsilon_{1}=\epsilon_{2}$, thus they are equal to $\epsilon$ only when two D-strings move so that their intersection points with the the third D-string are moving in the same direction at the light speed.}. In the conventional representation discussed above, these compatibility conditions result in the same relations among the coefficients as $a=d$ and $b=c$, therefore ensure $8$ supercharges. Hence we will have a $1/4$ supersymmetry for the generic case of three moving D-strings if their intersecting points form a triangle moving at the light speed without deforming its shape and size. Fig. \ref{figure5} shows the situation.

One might think of adding more D-strings to make general polygon shaped configurations. These additional D-strings will not perturb the supersymmetry as long as their corresponding compatibility conditions $\cos{\theta_{i}}\cos{\gamma_{i}}=1$ are satisfied. For $k$ D-strings, $k-1$ compatibility conditions are enough to show that the rest compatibility conditions are redundant. All these compatibility conditions result in the same relations among the coefficients (as $a=d$ and $b=c$), therefore ensure $1/4$ supersymmentry always. 

\FIGURE{
\epsfig{file=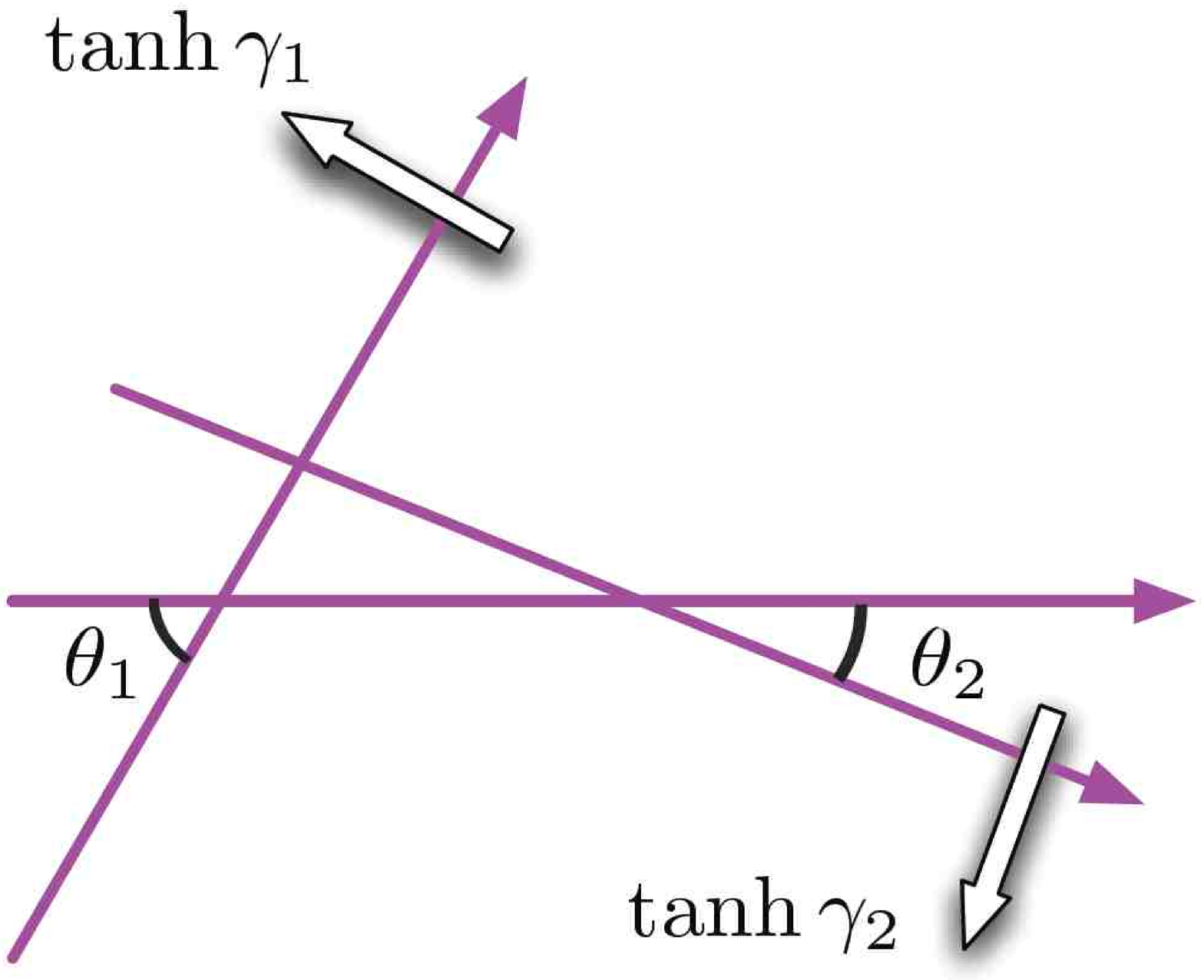,width=9cm} 
\caption{\small A supersymmetric triangle configuration.}
\label{figure5}}

%

\section{Two D-strings Moving in Different Planes}\label{secv}
Let the first D-string be tilted at angle $\theta_{1}$ and moving with rapidity $\gamma_{1}$ in $(x^{1},\,x^{2})$-plane and the second D-string be tilted at angle $\theta_{2}$ and moving with rapidity $\gamma_{2}$ in $(x^{3},\,x^{4})$-plane. 

The supercharges concerned with these strings are 
\begin{equation}\label{}
Q+\beta'^{2}\beta^{3}\beta^{4}\beta^{\bot}_{4}\tilde{Q}, \qquad Q+\beta'^{4}\beta^{1}\beta^{2}\beta^{\bot}_{4}\tilde{Q}
\end{equation}  
respectively. There, 
\begin{equation}\label{}
\beta'^{2}=\rho(-\theta_{1})\rho(-\gamma_{1})\beta^{2}\rho(\gamma_{1})\rho(\theta_{1}),\qquad\beta'^{4}=\rho(-\theta_{2})\rho(-\gamma_{2})\beta^{4}\rho(\gamma_{2})\rho(\theta_{2})
\end{equation}  
with
\begin{eqnarray}\label{}
&&\rho(\theta_{1})=e^{\Gamma^{12}\theta_{1}},\quad \rho(\gamma_{1})=e^{\Gamma^{02}\gamma_{1}},\nonumber\\
&&\rho(\theta_{2})=e^{\Gamma^{34}\theta_{2}},\quad \rho(\gamma_{2})=e^{\Gamma^{04}\gamma_{2}}.
\end{eqnarray}
The number of supercharges compatible with both D-strings is the number of the eigen spinors satisfying
\begin{eqnarray}\label{eig}
&&\left[(\beta^{4})^{-1}(\beta^{3})^{-1}\rho(-\theta_{1})\rho(-\gamma_{1})(\beta^{2})^{-1}\rho(\gamma_{1})\rho(\theta_{1})\rho(-\theta_{2})\rho(-\gamma_{2})\beta^{4}\rho(\gamma_{2})\rho(\theta_{2})\beta^{1}\beta^{2} \right] \epsilon\nonumber\\
&&=\left(-\Gamma^{0}\sinh{\gamma_{1}}- \Gamma^{1}\sin{\theta_{1}}\cosh{\gamma_{1}}+ \Gamma^{2}\cos{\theta_{1}}\cosh{\gamma_{1}} \right)\Gamma^{12} \nonumber\\
&&\qquad\cdot \left(\Gamma^{0}\sinh{\gamma_{2}}- \Gamma^{3}\sin{\theta_{2}}\cosh{\gamma_{2}}+ \Gamma^{4}\cos{\theta_{2}}\cosh{\gamma_{2}} \right) \Gamma^{34}\epsilon \nonumber\\
&&=\epsilon
\end{eqnarray}
 In $\mathbf{s}$-representation,
 \begin{eqnarray}\label{}
\epsilon&=&a_{+}|+,+,+,s^{3},s^{4}>+b_{+}|+,-,+,s^{3},s^{4}>\nonumber\\
&&+c_{+}|-,+,+,s^{3},s^{4}>+d_{+}|-,-,+,s^{3},s^{4}>\nonumber\\
 &&+a_{-}|+,+,-,s^{3},s^{4}>+b_{-}|+,-,-,s^{3},s^{4}>\nonumber\\
 &&+c_{-}|-,+,-,s^{3},s^{4}>+d_{-}|-,-,-,s^{3},s^{4}>\nonumber\\
 &\equiv&(a_{+},b_{+},c_{+},d_{+};a_{-},b_{-},c_{-},d_{-}),
\end{eqnarray}  
 we note that
 \begin{eqnarray}\label{}
\Gamma^{1234}\epsilon&=&(-a_{+},b_{+},-c_{+},d_{+};a_{-},-b_{-},c_{-},-d_{-}), \nonumber\\
\Gamma^{0124}\epsilon&=&(-c_{-},-d_{-}, -a_{-},-b_{-};c_{+},d_{+},a_{+},b_{+}), \nonumber\\
\Gamma^{0123}\epsilon&=&-i(c_{-},d_{-}, a_{-},b_{-};c_{+},d_{+},a_{+},b_{+}),\nonumber\\
\Gamma^{0234}\epsilon&=&(d_{+},-c_{+},b_{+},-a_{+};-d_{-},c_{-},-b_{-},a_{-}), \nonumber\\
\Gamma^{0134}\epsilon&=&i(d_{+},c_{+},b_{+},a_{+};-d_{-},-c_{-},-b_{-},-a_{-}), \nonumber\\
\Gamma^{24}\epsilon&=&(-b_{-},-a_{-},-d_{-},-c_{-};b_{+},a_{+},d_{+},c_{+}), \nonumber\\
\Gamma^{23}\epsilon&=&-i(b_{-},a_{-},d_{-},c_{-};b_{+},a_{+},d_{+},c_{+}), \nonumber\\
\Gamma^{14}\epsilon&=&i(-b_{-},a_{-},-d_{-},c_{-};b_{+},-a_{+},d_{+},-c_{+}), \nonumber\\
\Gamma^{13}\epsilon&=&(b_{-},-a_{-},d_{-},-c_{-};b_{+},-a_{+},d_{+},-c_{+}).
\end{eqnarray}
With the insertion of these results, the eigen spinor equation (\ref{eig}) has nontrivial solutions only if its characteristic equation vanishes, that is, $4\cosh^{2}{\gamma_{1}}\cosh^{2}{\gamma_{2}}=0$, but this is impossible. Hence there is no supersymmetry preserved in the configuration of two D-strings moving in different planes which are not intersecting.

For lateral use, we add a few more computational results of gamma matrices acting the spinors;
\begin{eqnarray}\label{}
\Gamma^{01}\epsilon&=&(d_{+},c_{+},b_{+},a_{+};d_{-},c_{-},b_{-},a_{-}), \nonumber\\
\Gamma^{02}\epsilon&=&i(-d_{+},c_{+},-b_{+},a_{+};-d_{-},c_{-},-b_{-},a_{-}), \nonumber\\
\Gamma^{03}\epsilon&=&(-c_{-},d_{-},-a_{-},b_{-};-c_{+},d_{+},-a_{+},b_{+}), \nonumber\\
\Gamma^{12}\epsilon&=&i(a_{+},-b_{+},c_{+},-d_{+};a_{-},-b_{-},c_{-},-d_{-}), \nonumber\\
\Gamma^{34}\epsilon&=&i(a_{+},b_{+},c_{+},d_{+};-a_{-},-b_{-},-c_{-},-d_{-}).
\end{eqnarray}

\section{Supersymmetric D-Jackstraws}\label{secvi}
Let us generalize the scissors configuration to higher dimensions. We consider three D-strings each of which are moving but not necessarily in the same plane. Let the first D-string be tilted at angle $\theta_{1}$ and move in $(x^{1},x^{2})$-plane and with a rapidity $\gamma_{1}$. The second D-string is tilted at an angle $\theta_{2}$ and is moving in $(x^{2},x^{3})$-plane with a rapidity $\gamma_{2}$. We align one D-string along $x^{2}$-direction. See Fig. \ref{figure6} for the configuration. 

\FIGURE{
\epsfig{file=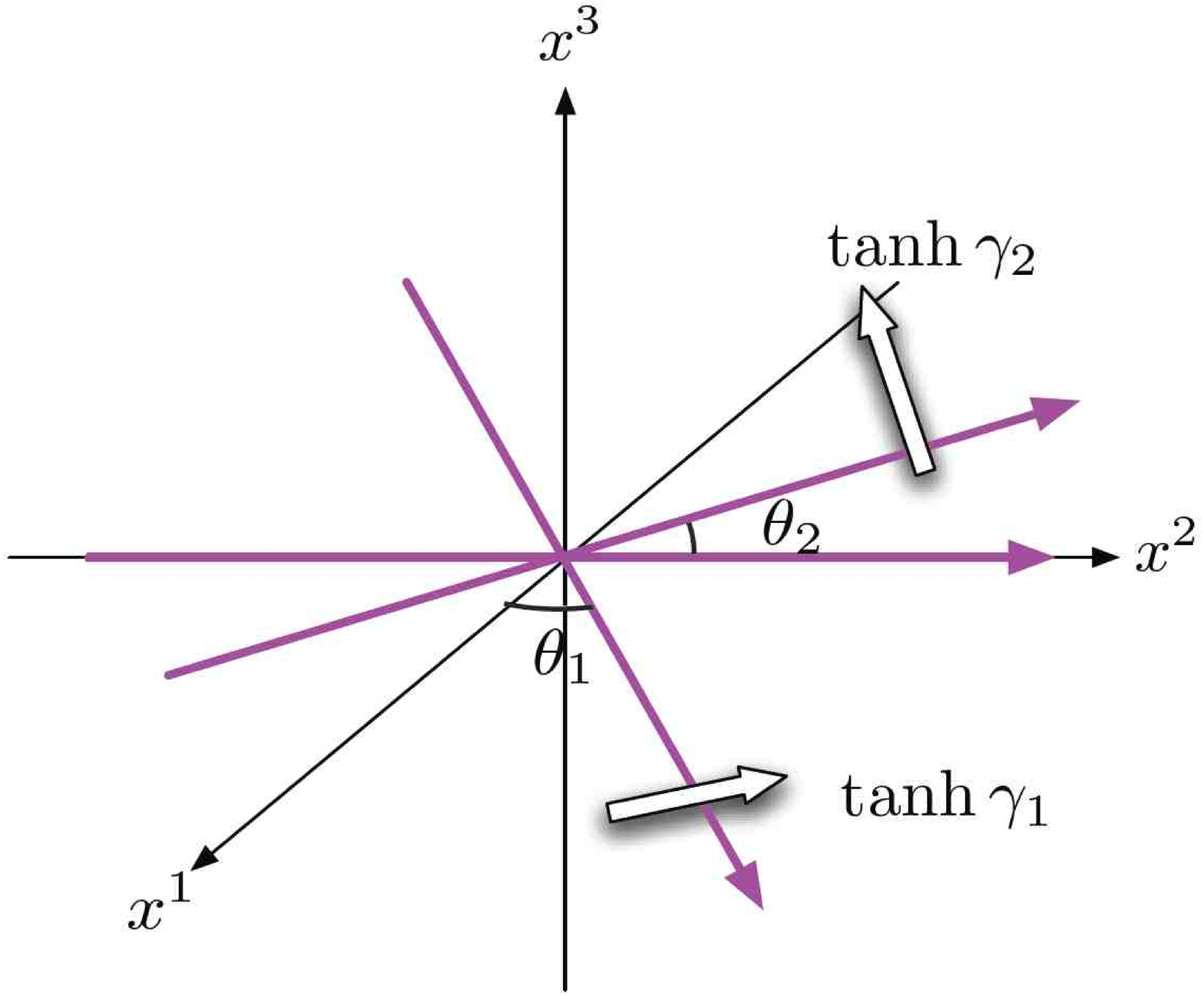,width=9cm} 
\caption{\small A non-planar array of three intersecting D-strings: The first D-string is moving in $(x^{3},x^{1})$-plane while the second one is moving in $(x^{2},x^{3})$-plane. The third D-string is along $x^{2}$-axis.}
\label{figure6}}

%

The supercharges concerning each of these D-strings are
\begin{equation}\label{supercharge3}
Q+\beta'^{2}\beta^{3}\beta^{\bot}_{3}\tilde{Q},\quad
Q+\beta'^{3}\beta^{1}\beta^{\bot}_{3}\tilde{Q},\quad
Q+\beta^{3}\beta^{1}\beta^{\bot}_{3}\tilde{Q}
\end{equation}
respectively.
The number of supercharges compatible with the whole configuration is the number of spinors $\epsilon$ satisfying
\begin{eqnarray}\label{conds}
\left( \beta^{1} \right)^{-1}\beta'^{2}\epsilon&=&\Gamma^{12}\rho(\theta_{1})\rho(2\gamma_{1})\rho(\theta_{1})\epsilon\\
&=&\left[\cosh{\gamma_{1}}\left(\Gamma^{12}\cos{\theta_{1}}-\sin{\theta_{1}} \right) +\Gamma^{01}\sinh{\gamma_{1}} \right]\epsilon=-\epsilon, \nonumber\\
\left(\beta^{3} \right)^{-1}\beta'^{3}\epsilon&=&\rho(\theta_{2})\rho(2\gamma_{2})\rho(\theta_{2})\epsilon=\left[\cosh{\gamma_{2}} \left( \cos{\theta_{2}}+\Gamma^{23} \sin{\theta_{2}} \right) +\Gamma^{03} \sinh{\gamma_{2}}	 \right] \epsilon=\epsilon,\nonumber
\end{eqnarray}
where $\beta'^{2}=\rho(-\theta_{1})\rho(-\gamma_{1})\beta^{2}\rho(\gamma_{1})\rho(\theta_{1})$ and $\beta'^{3}=\rho(-\theta_{2})\rho(-\gamma_{2})\beta^{3}\rho(\gamma_{2})\rho(\theta_{2})$. These equations are the compatibility condition between the first and the third, and between the second and the third D-string respectively.  

In $\mathbf{s}$-representation discussed in the previous sections, the explicit form of the conditions (\ref{conds}) are
\begin{eqnarray}\label{}
&&\left(\sin{\theta_{1}}\cosh{\gamma_{1}}-1 \right)(a_{+},b_{+},c_{+},d_{+};a_{-},b_{-},c_{-},d_{-})\nonumber\\
 &&-i \cos{\theta_{1}}\cosh{\gamma_{1}}(a_{+},-b_{+},c_{+},-d_{+};a_{-},-b_{-},c_{-},-d_{-}) \nonumber\\
&&- \sinh{\gamma_{1}}(d_{+},c_{+},b_{+},a_{+};d_{-},c_{-},b_{-},a_{-})=0, \nonumber\\&& \nonumber\\
&&\left(\cos{\theta_{2}}\cosh{\gamma_{2}}-1 \right)(a_{+},b_{+},c_{+},d_{+};a_{-},b_{-},c_{-},d_{-})\nonumber\\ 
&&-i \sin{\theta_{2}}\cosh{\gamma_{2}} (b_{-},a_{-},d_{-},c_{-};b_{+},a_{+},d_{+},c_{+}) \nonumber\\
&&+ \sinh{\gamma_{2}}(-c_{-},d_{-},-a_{-},b_{-};-c_{+},d_{+},-a_{+},b_{+})=0.
 \end{eqnarray}
These two equations have nontrivial solutions if their characteristic equations are satisfied;
\begin{equation}\label{}
\sin{\theta_{1}}\cosh{\gamma_{1}}-1=0,\qquad \cos{\theta_{2}}\cosh{\gamma_{2}}-1=0.
\end{equation} 
These conditions imply that $0<\theta_{1}<\pi$ and $-\pi/2<\theta_{2}<\pi/2$, therefore the intersecting angle with the third D-string lying along $x^{2}$-direction should be less than $\pi/2$. The conditions do not constrain the motion of two D-strings completely and there are ambiguities in their moving directions as $\cos{\theta_{1}}=p\,\tanh{\gamma_{1}}$ and $\sin{\theta_{2}}=q\,\tanh{\gamma_{2}}$ with $p^{2}=q^{2}=1$. With either sign of $p$ and $q$, two D-strings slide over the third static D-string at the light speed. However, in order for these two D-strings to be compatible with each other preserving supersymmetries, the ambiguities $p$ and $q$ should be further constrained so that $p\,q <0$. This is when their intersection points (with the third D-string) move in the same direction along the third D-string.
For example let $0<\theta_{1,2}<\pi/2$. Then $\cos{\theta_{1}}=\tanh{\gamma_{1}}$ with $\gamma_{1}>0$ and $\sin{\theta_{2}}=-\tanh{\gamma_{2}}$ with $\gamma_{2}<0$ and the above two equations consistently result in the relations among the coefficients;
\begin{equation}\label{relat}
i a_{\pm}=-d_{\pm},\qquad
ib_{\pm}=c_{\pm}
\end{equation} 
leaving $8$ supersymmetries. 

One can consider more general situation where the first D-string moves in $(x'^{1},x^{2})$-plane where $(x'^{3},x'^{1})$-axes are rotated counterclockwise by an angle $\phi$ with respect to $(x^{3},x^{1})$-axes. Therefore the first and the second D-strings are moving in two planes which are intersecting at an angle $\pi/2-\phi$. Fig. \ref{figure7} exhibits the configuration.

\FIGURE{
\epsfig{file=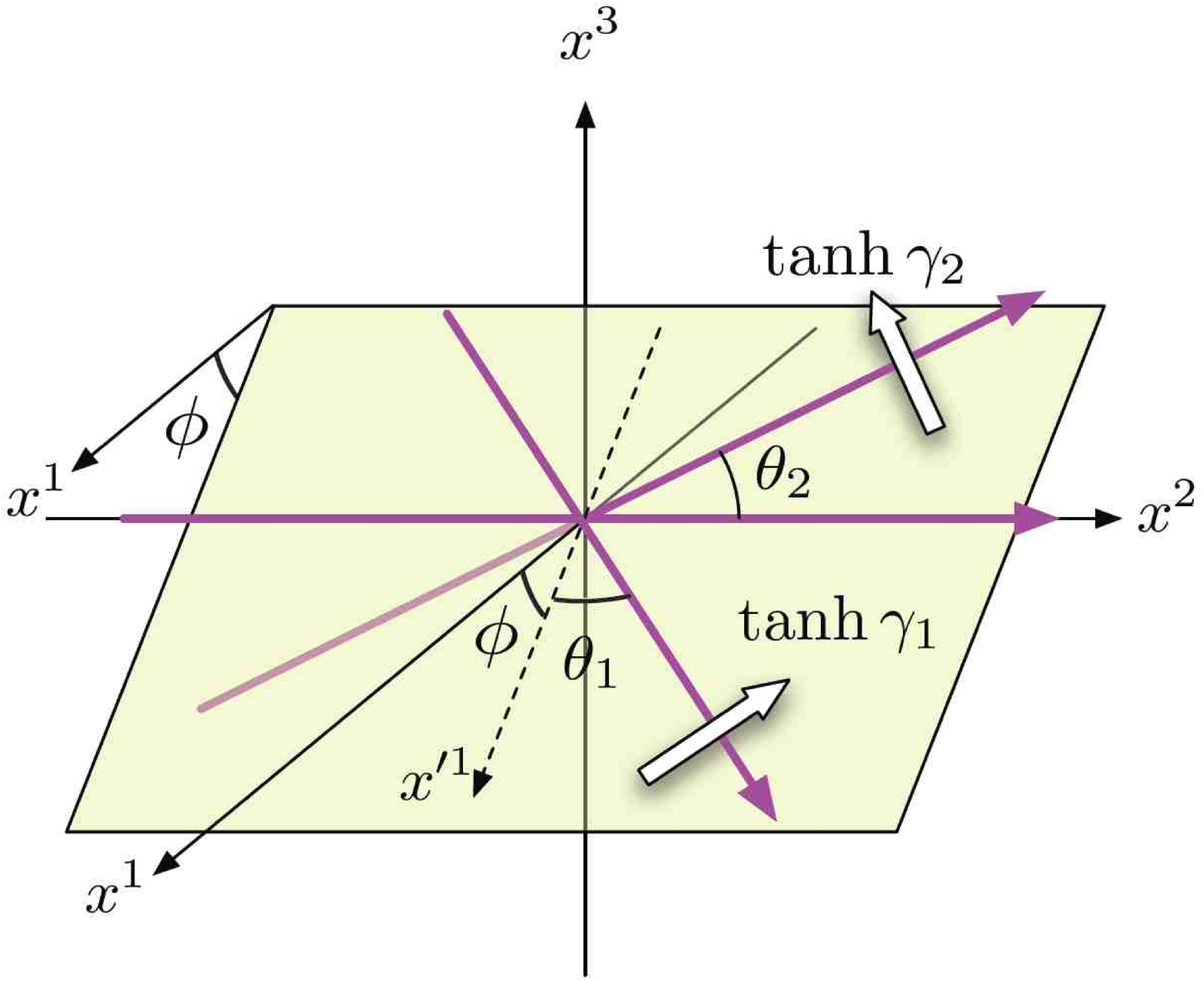,width=7cm} 
\caption{\small A configuration of three intersecting D-strings. The first is moving in $(x'^{1},x^{2})$-plane while the second one is moving in $(x^{2},x^{3})$-plane. The third D-string is along $x^{2}$-axis.}
\label{figure7}
}

%

One can make an educated guess that this new configuration, interpolating the the previous one and the triangular one, will preserve $8$-supercharges too. This is indeed the case. The difference caused by this change appears in the supercharge concerning the first D-string so that the first equation of (\ref{supercharge3}) becomes
\begin{equation}\label{}
Q+\rho(\phi)\beta'^{2}\beta^{3}\rho(-\phi)\beta^{\bot}_{3}\tilde{Q},
\end{equation}  
where $\rho(\phi)= \cos{(\phi/2)}+\Gamma^{31}\sin{(\phi/2)}$. The story goes in the same way as in the previous case and the first compatibility condition of Eq (\ref{conds}) is changed to
\begin{eqnarray}\label{}
&&\left(\beta^{3} \right)^{-1} \left(\beta^{1} \right)^{-1} \rho(\phi)\beta'^{2}\beta^{3}\rho(-\phi)\epsilon\nonumber\\
&&=\left[\cos{\phi} \left( \cos{\theta_{1}}\cosh{\gamma_{1}}+\Gamma^{02}\sinh{\gamma_{1}}\right)+ \Gamma^{12}\sin{\theta_{1}}\cosh{\gamma_{1}}\right. \nonumber\\
&&\left. \qquad- \Gamma^{31}\sin{\phi}\left(\cos{\theta_{1}}\cosh{\gamma_{1}}+\Gamma^{02}\sinh{\gamma_{1}} \right)  \right]  \epsilon\nonumber\\
&&=\epsilon.
\end{eqnarray}
In $\mathbf{s}$-representation, this condition is specified as
\begin{eqnarray}\label{newcond}
&& \left(\sin{\theta_{1}}\cosh{\gamma_{1}}-1 \right)(a_{+},b_{+},c_{+},d_{+};a_{-},b_{-},c_{-},d_{-})\nonumber\\
 &&-i \cos{\phi}\cos{\theta_{1}}\cosh{\gamma_{1}}(a_{+},-b_{+},c_{+},-d_{+};a_{-},-b_{-},c_{-},-d_{-}) \nonumber\\
&&-\cos{\phi} \sinh{\gamma_{1}}(d_{+},c_{+},b_{+},a_{+};d_{-},c_{-},b_{-},a_{-}) \nonumber\\
 &&+i \sin{\phi}\cos{\theta_{1}}\cosh{\gamma_{1}}(b_{-},a_{-},d_{-},c_{-};b_{+},a_{+},d_{+},c_{+}) \nonumber\\
&&+\sin{\phi} \sinh{\gamma_{1}}(-c_{-},d_{-},-a_{-},b_{-};-c_{+},d_{+},-a_{+},b_{+})=0.
\end{eqnarray}
In order for this equation to have non-trivial solutions for the coefficients, the following characteristic equation should be satisfied;
\begin{equation}\label{}
\left(2-2 \sin{\theta_{1}}\cosh{\gamma_{1}} \right)^{2}=0. 
\end{equation}
Since this equation is independent of the angle $\phi$ and the new condition (\ref{newcond}) gives then the same relations as (\ref{relat}), we conclude that this more generic case of three moving D-strings preserves $8$ supercharges. 

Summing up all the results so far, we can say that an arbitrary number of D-strings, intersecting at an arbitrary angle less than $\pi$ with one another and moving at an arbitrary rapidity, preserve a quarter supersymmetry so long as the spatial trajectories of D-strings have one common direction and all the intersection points are moving at the light speed without deforming their relative configuration. See Fig. \ref{random} for an example of supersymmetric network of moving D-strings in $3$-dimensions. If all the intersection points of D-strings coincide, the whole configuration will compose a null hedgehog that preserves $8$ supercharges.  

\FIGURE{
\epsfig{file=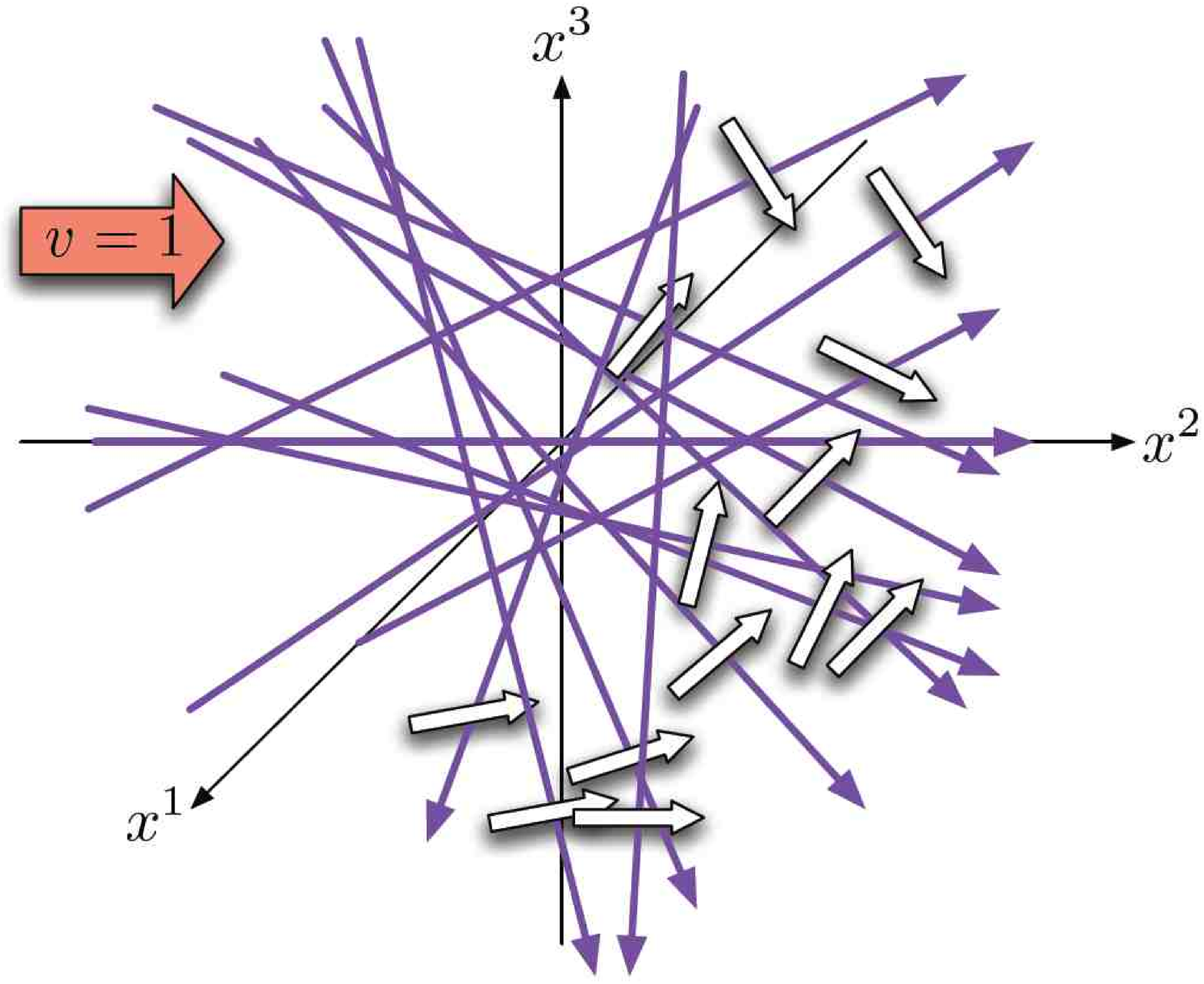,width=7cm}
\epsfig{file=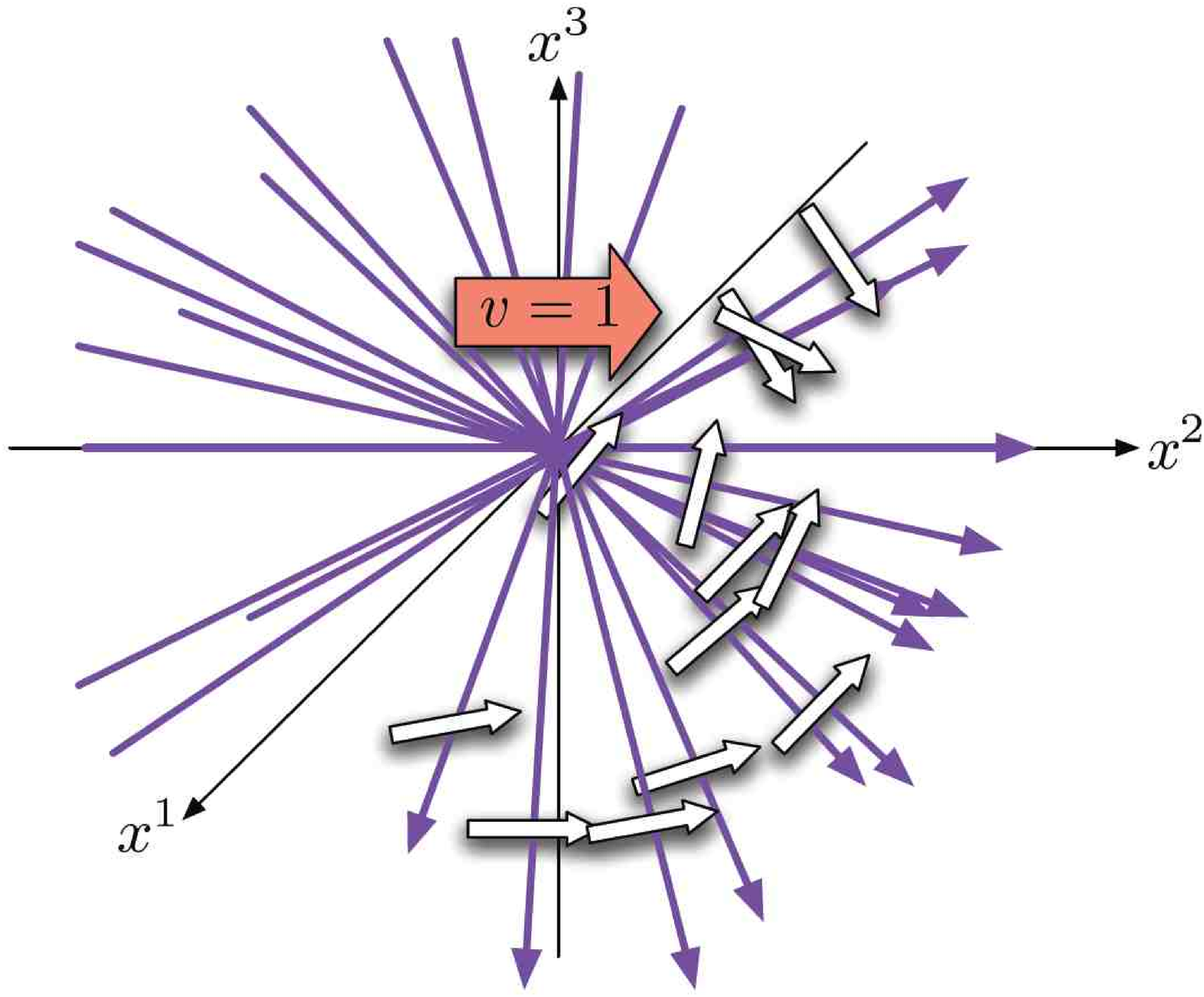,width=7cm} 
\caption{\small The left figure shows a supersymmetric arrangement of moving D-strings (that looks like Jackstraws). All the D-strings are moving in such a way that all the intersection points move in one direction at the light speed keeping  their relative positions. The right figure: With all the intersection points coincident, the configuration becomes a super null hedgehog.}
\label{random}
}

%

\section{Super Waterwheels}\label{secvii}
Now let us take T-duality on the super hedgehog solutions which have been discussed so far. Instead of dealing with the whole configuration, we first focus on one D-string and see the effect of T-duality on it. Let the D-string lie in the $(x^{1},x^{2})$-plane, be tilted at an angle $\theta$ with respect to $x^{1}$-axis, and be boosted at a rapidity $\gamma$. The boundary conditions for the open string ending on the D-string are
\begin{eqnarray}\label{}
&&\partial_{\sigma}x^{0}|\cosh{\gamma}+\partial_{\sigma}x^{1}\sin{\theta}\sinh{\gamma}-\partial_{\sigma}x^{2}|\cos{\theta}\sinh{\gamma}=0 \nonumber\\
&&\partial_{\sigma}x^{1}|\cos{\theta}+\partial_{\sigma}x^{2}|\sin{\theta}=0 \nonumber\\
&&-\partial_{\tau}x^{0}|\sinh{\gamma}-\partial_{\tau}x^{1}|\sin{\theta}\cosh{\gamma}+\partial_{\tau}x^{2}|\cos{\theta}\cosh{\gamma}=0.
\end{eqnarray}
where the vertical bar `$|$' is short for `$|_{\sigma=0,\pi}$' that means the quantity in front of this bar measured at the world-sheet boundary $\sigma=0,\pi$. Taking T-duality along $x^{2}$-axis interchanges the term $\partial_{\sigma}x^{2}$ with $\partial_{\tau}x^{2}$ in the boundary conditions and vice versa. The boundary conditions of the string living on T-dualized configuration are
\begin{eqnarray}\label{boundary1}
&&\partial_{\sigma}x^{0}|\cosh{\gamma}+\partial_{\sigma}x^{1}\sin{\theta}\sinh{\gamma}-\partial_{\tau}x^{2}|\cos{\theta}\sinh{\gamma}=0 \nonumber\\
&&\partial_{\sigma}x^{1}|\cos{\theta}+\partial_{\tau}x^{2}|\sin{\theta}=0 \nonumber\\
&&-\partial_{\tau}x^{0}|\sinh{\gamma}-\partial_{\tau}x^{1}|\sin{\theta}\cosh{\gamma}+\partial_{\sigma}x^{2}|\cos{\theta}\cosh{\gamma}=0.
\end{eqnarray}

Comparing these conditions with that of a string coupled to the background antisymmetric field $\mathcal{F}_{\mu\nu}$
\begin{equation}\label{}
 \left( G_{\mu\nu}\partial_{\sigma}x^{\nu}-\mathcal{F}_{\mu\nu}\partial_{\tau}x^{\nu}\right)|_{\sigma=0,\pi} =0
\end{equation}
we note that the string lives on a D$2$-brane containing background fields over its world volume as
\begin{equation}\label{}
\mathcal{F}_{02}=- \frac{\tanh{\gamma}}{\cos{\theta}},\qquad \mathcal{F}_{12}=-\tan{\theta}.
\end{equation}       

Now we consider $N$ D-strings, each of which is moving in $(x^{\mu_{a}},x^{2})$-plane at a rapidity $\gamma_{a}$, being tilted at an angle $\theta_{a}$ with respect to $x^{\mu_{a}}$-axis. Here $a=1,\,2,\cdots,N-1$ and they need not orthogonal to one another. Without loss of generality, we let the last D-string lie along $x^{2}$-axis without moving. Taking T-duality along $x^{2}$-axis results in a complex of $N-1$ D$2$-branes, each of which is extending in $(x^{\mu_{a}},x^{2})$-plane and contains the electric field and the magnetic field as
\begin{equation}\label{}
\mathcal{F}^{(a)}_{02}=- \frac{\tanh{\gamma_{a}}}{\cos{\theta_{a}}},\qquad \mathcal{F}^{(a)}_{\mu_{a}2}=-\tan{\theta_{a}},
\end{equation}   
and an array of D$0$-branes smeared over $x^{2}$-axis. If we start with a supersymmetric null hedgehog solution so that the parameters $\theta_{a}$ and $\gamma_{a}$ satisfy the conditions
\begin{equation}\label{}
\sin{\theta_{a}}\cosh{\gamma_{a}}-1=0,\qquad (a=1,\,2,\cdots, N-1),
\end{equation}    
all the electric fields become `critical', i.e., $E\equiv\mathcal{F}^{(a)}_{20}=\pm 1$. From here on, we choose it as $E=1$, to simplify the argument. Fig. \ref{figure8} shows the resulting configuration.

\FIGURE{
\epsfig{file=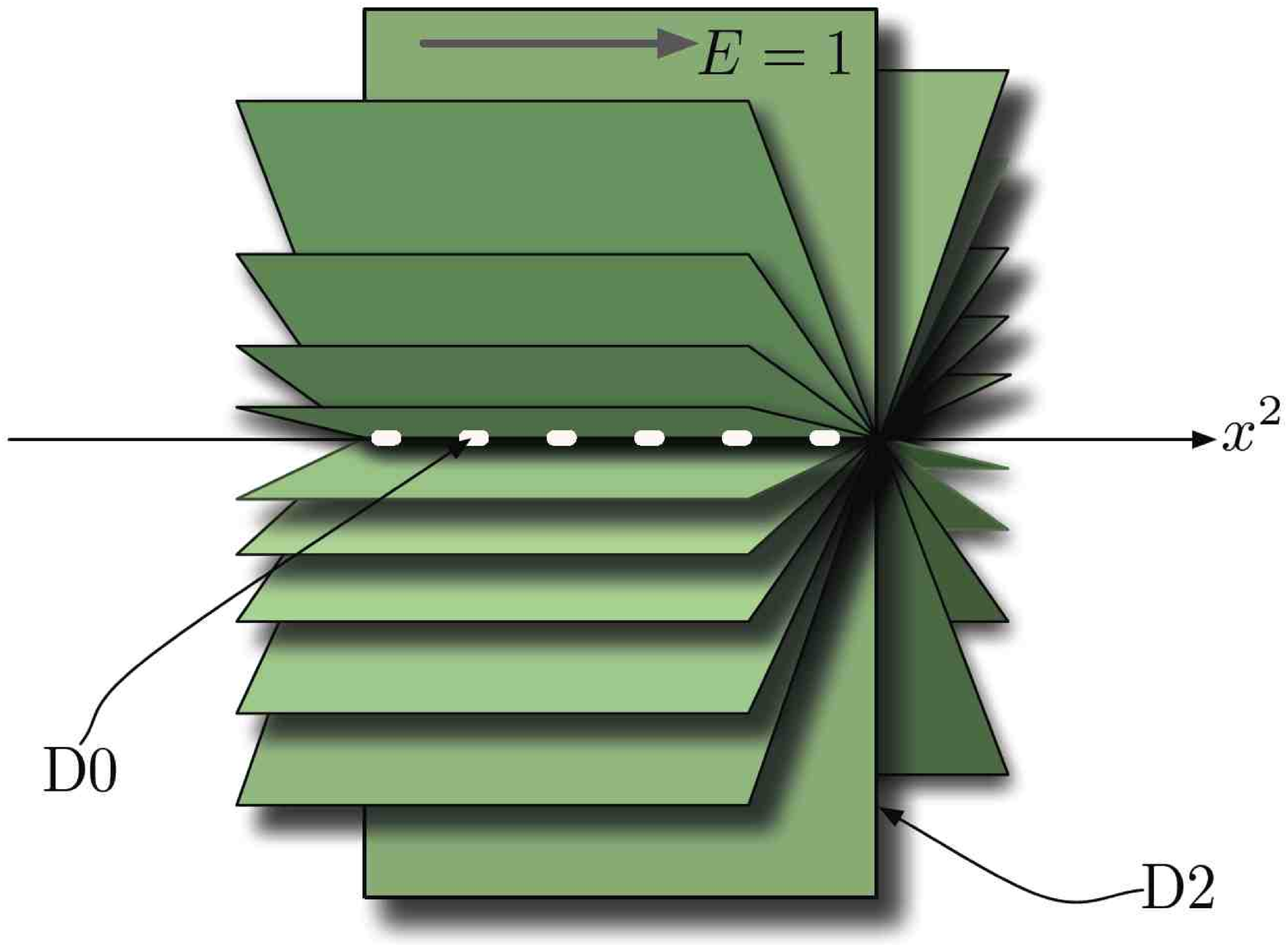,width=9cm} 
\caption{\small A supersymmetric threshold bound state of intersecting D$2$-branes and an array of D$0$-branes (white dots) smeared over $x^{2}$-axis. Only two spatial directions transverse to the D$0$ array are shown with other directions omitted.}
\label{figure8}}

%

Especially had we allowed the momentum flow on the last D-string so that it make a super D-helix, the type IIA configuration obtained via T-duality would have been a supertube with $N-1$ wings. It is surprising to see that the whole configuration preserves $8$-supercharges regardless of the intersecting angles among the wings.

\section{Supersymmetric $(p,\,q)$-Jackstraws}\label{secviii}
In this section, we display a planar configuration composed of $(p,\,q)$-strings preserving $8$ supersymmetries. We start with a D$2$-brane with Born-Infeld fields,
\begin{equation}\label{}
\mathcal{F}_{02}=-\frac{\tanh{\gamma}}{\cos{\theta}},\qquad \mathcal{F}_{12}=-\tan{\theta}.
\end{equation}
By taking T-duality along $x^{1}$-direction, we get a D-string tilted at an angle $\theta$ with respect to $x^{2}$-axis and with Born-Infeld electric field along its volume. This can be easily seen from the corresponding boundary conditions of the open string living on the D-string. T-duality along $x^{1}$-axis results in the change $\partial_{\tau}x^{1}\leftrightarrow\partial_{\sigma}x^{1}$ in the boundary condition (\ref{boundary1}), so leads to
\begin{eqnarray}\label{boundary2}
&&\partial_{\sigma}x^{0}|\cosh{\gamma}+\partial_{\tau}x^{1}\sin{\theta}\sinh{\gamma}-\partial_{\tau}x^{2}|\cos{\theta}\sinh{\gamma}=0, \nonumber\\
&&\partial_{\tau}x^{1}|\cos{\theta}+\partial_{\tau}x^{2}|\sin{\theta}=0, \nonumber\\
&-&\partial_{\tau}x^{0}|\sinh{\gamma}-\partial_{\sigma}x^{1}|\sin{\theta}\cosh{\gamma}+\partial_{\sigma}x^{2}|\cos{\theta}\cosh{\gamma}=0.
\end{eqnarray}
Introducing new coordinates $\bar{x}^{0}=x^{0}$, $\bar{x}^{1}=x^{1}\cos{\theta}+x^{2}\sin{\theta}$ and $\bar{x}^{2}=x^{2}\cos{\theta}-x^{1}\sin{\theta}$, that is, the coordinates rotated counterclockwise by an angle $\theta$ with respect to the original coordinates $\{x^{0,1,2}\}$, we can write the boundary conditions as
\begin{eqnarray}\label{}
&&\partial_{\sigma}\bar{x}^{0}|\cosh{\gamma}-\partial_{\tau}\bar{x}^{2}|\sinh{\gamma}=0, \nonumber\\
&&\partial_{\tau}\bar{x}^{1}|=0, \nonumber\\
&-&\partial_{\tau}\bar{x}^{0}|\sinh{\gamma}+\partial_{\sigma}\bar{x}^{2}|\cosh{\gamma}=0.
\end{eqnarray}
Hence $\bar{x}^{1}$ is one of Dirichlet directions and the D-string extends to the direction $\bar{x}^{2}$ with Born-Infeld electric field 
\begin{equation}\label{}
\mathcal{F}_{\bar{0}\bar{2}}=-\tanh{\gamma}.
\end{equation} 
 Note that the direction $\bar{x}^{2}$ is rotated counterclockwise by an angle $\theta$ with respect to $x^{2}$-axis. 
 
 Had we started with $N$ D$2$-branes superposed but with different Born-Infeld fields as $\mathcal{F}^{(a)}_{02}=-\tanh{\gamma_{(a)}}/\cos{\theta_{(a)}}$ and $\mathcal{F}^{(a)}_{12}=-\tan{\theta_{(a)}}$ ($a=1,\,2,\cdots, N$), we would have got $N$ D-strings, each of which is tilted at an angle $\theta_{(a)}$ with respect to $x^{2}$-axis and has Born-Infeld electric field of magnitude $\mathcal{F}^{(a)}_{\bar{0}\bar{2}}=-\tanh{\gamma_{(a)}}$ along its direction. 
 
 Especially when all the conditions $\sin{\theta_{(a)}}\cosh{\gamma_{(a)}}=1$ are satisfied, the whole planar configuration of static gauged D-strings preserves $8$ supercharges. In the case, the tilting angles $\theta_{(a)}$ of the gauged D-strings are correlated with the magnitude of BI fields on their world volumes: 
\begin{equation}\label{}
E_{(a)}=-\mathcal{F}^{(a)}_{\bar{0}\bar{2}}=\tanh{\gamma_{(a)}}=\cos{\theta_{(a)}}.
\end{equation} 
This implies that, in order not to break any supersymmetry, we pose a $(p_{(a)},\,1)$-string with 
\begin{equation}\label{p-charge}
p_{(a)}= \frac{E_{(a)}}{\lambda\sqrt{1-E^{2}_{(a)}}}=\frac{1}{\lambda}\cot{\theta_{(a)}}
\end{equation}   
at the angle $\theta_{(a)}$ (counted counterclockwise from $x^{2}$-axis). 

In general, the angular position of a specific $(\bar{p},\,\bar{q})$-string, in the planar configuration of Jackstraws, is determined by its charge ratio as
\begin{equation}\label{angle}
\frac{\bar{q}}{\bar{p}}=\lambda\tan{\theta}.
\end{equation}  
This is based the fact that multiple $(p_{(a)},\,1)$-strings can be posed in parallel or be superposed on top of each other without breaking any supersymmetry. The angular position of a $(n\,p_{(a)},\,n)$-string, i.e., $n$ multiples of a $(p_{(a)},\,1)$-string, is independent of $n$. A special orthogonally intersecting case of a D-string and a F-string was discussed in the name of `super cross' in Ref. \cite{Chen:2003ir}.

The formula (\ref{angle}) is reminiscent of the S-dual transformation that leads to a $(\bar{p},\,\bar{q})$-string, being acted on a $(1,\,0)$-string. Indeed the angle coincides with that appearing in $SO(2)$ subgroup of $SL(2,\,\mathbf{R})$, the group of S-dual transformations. We first note that the angle $\theta$ was measured counterclockwise from $x^{2}$-axis, that is, the direction of a $(\infty,\,1)$-string or nearly that of a $(1,\,0)$-string. In the absence of the R-R scalar field concerning D-instanton, the required S-dual transformation is the form,
\begin{eqnarray}\label{superangle}
\frac{1}{\sqrt{\bar{p}^{2}+\bar{q}^{2}}}\left(\begin{array}{c}\bar{p} \\\bar{q}\end{array}\right)_{\lambda}
&=& \left(\begin{array}{cc}1/\sqrt{\lambda} & 0 \\0 & \sqrt{\lambda}\end{array}\right)
\left(\begin{array}{rr}\cos{\theta} &\quad -\sin{\theta} \\\sin{\theta} & \cos{\theta}\end{array}\right)
\left(\begin{array}{c}1 \\0\end{array}\right)_{1}.
\end{eqnarray}
The first factor on the right hand side concerns the non-trivial string coupling, $\lambda$, and the next factor transforms the charge $(1,\,0)_{1}$ to $(\cos{\theta},\,\sin{\theta})_{1}$. The subscript in each charge doublet denotes the value of the string coupling constant characterizing the vacuum. We introduced the factor $\sqrt{\bar{p}^{2}+\bar{q}^{2}}$ to account for the integer values, $\bar{p}$ and $\bar{q}$. See Ref. \cite{Schwarz:1995dk} for details.

One need not refer to a fundamental string, or  $x^{2}$-axis, as the only angle basis to use in making the super Jackstraws. One can easily change the reference string to another (a $(p,\,q)$-string for example) by making use of the above relation (\ref{superangle}). Conclusively to say, the way to win the super $(p,\,q)$-Jackstraws game is as follows: When we are given $(p,\,q)$-string and want to pose a $(p',\,q')$-string without breaking supersymmetry, we tune the intersecting angle $\theta$ so that 
\begin{eqnarray}\label{}
\frac{1}{\sqrt{p'^{2}+q'^{2}}}\left(\begin{array}{c}p' \\q'\end{array}\right)_{\lambda}&=&\frac{1}{\sqrt{p^{2}+q^{2}}}\left(\begin{array}{rr}\cos{\theta} &\quad -\lambda^{-1}\sin{\theta}\\\lambda\sin{\theta} & \cos{\theta}\end{array}\right)\left(\begin{array}{c}p \\q\end{array}\right)_{\lambda}.
\end{eqnarray}   
These $(p,\,q)$-Jackstraws are related to the planar dynamic D-Jackstraws via T-dualities. Fig. \ref{figure9} illustrates the situation.

\FIGURE{
\epsfig{file=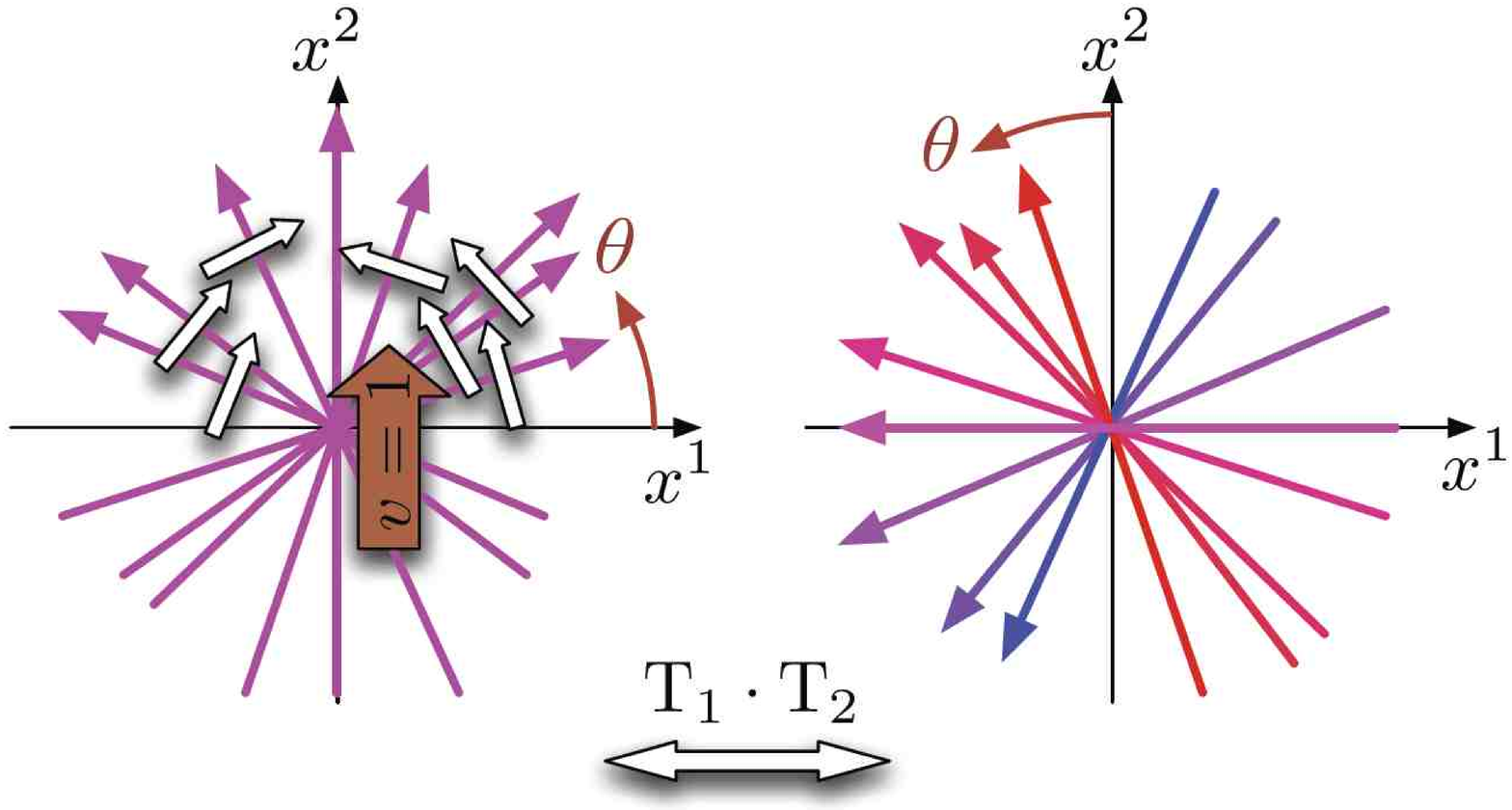,width=12cm} 
\caption{\small T-duality relates a planar null super D-hedgehog (on the left) with a planar static super $(p,\,q)$-hedgehog (on the right). The whole configuration is rotated by $\pi/2$ with respect to each other. Supersymmetry condition relates the rapidity and the tilting angle from $x^{1}$-axis (in the left figure) or BI electric field and the tilting angle from $x^{2}$ axis (in the right figure) as $\tanh{\gamma}=\cos{\theta}$.}
\label{figure9}}


\section{Discussions}\label{secix}

So far, we have considered various new configurations of D-branes preserving $8$ supersymmetries.  We conclude the paper with some remarks on the properties of those configurations. 

The intersecting angles among variously charged $(p,\,q)$-strings should be discrete at the quantum level. The classical S-duality group $SL(2,\,\mathbf{R})$ becomes $SL(2,\,\mathbf{Z})$ beyond the tree level. Basically this is due to the generalized Dirac quantization conditions valid in the presence of the magnetic dual partners of F-strings and D-strings, i.e., NS$5$-branes and D$5$-branes respectively. The angle in (\ref{angle}) is constraint by the discrete values of $p$ and $q$. 

Another interesting thing is that the angle depends on the string coupling constant. Given the charge pair $(p,\,q)$, the corresponding string approaches to the reference $x^{2}$-axis in the strong coupling limit, while it approaches to D-strings extending to the negative $x^{1}$-direction in the weak coupling limit.

In the super D-helix configuration \cite{Cho:2001ys}, one can understand its preserving $8$ supersymmetries in the differential view point. At every point along the helix, we consider the tangent line, that is nothing but a moving D-string. For arbitrary set of the points on the helix, the tangential D-strings form super D-Jackstraws. In the same vein, the D$2$-branes composing super waterwheels can be understood as the tangential planes to the supertubes. This view point is consistent with the arbitrary shape of the super D-helix \cite{Lunin:2001fv}\cite{Mateos:2002yf} or of the supertube \cite{Bak:2001xx}\cite{Mateos:2001pi}\cite{Hyakutake:2002fk}. In the super D-Jackstraws, we are free to move each constituent D-string as long as we keep its pointing direction. Therefore for a super D-helix of arbitrary shape, the tangential D-strings compose super D-Jackstraws. The same is true for the tangential D$2$-branes of a supertube of generic shape.   

The supersymmetry ensures the stability of the corresponding configuration. Let us consider the `joining and splitting interaction' in the ordinary intersecting D-strings. Unless two D-strings are in parallel, the joining and splitting procedure renders the configuration unstable. (See Ref. \cite{polchinski} for details.)
However, this interaction cannot happen in the intersection of two different $(p,\,q)$-strings because it will violate the charge conservation. In the spectrum of open strings extending over two different $(p,\,q)$-strings, there will be no tachyonic mode as long as $(p,\,q)$-strings tune their intersecting angle in the supersymmetric manner discussed so far. Actually the supersymmetry could give some new ingredients to D-, F-, or $(p,\,q)$-strings which are considered recently as the candidate cosmic strings \cite{Jones:2002cv}\cite{Sarangi:2002yt}. (See also Ref. \cite{Polchinski:2004ia} and references therein.)

\FIGURE{
\epsfig{file=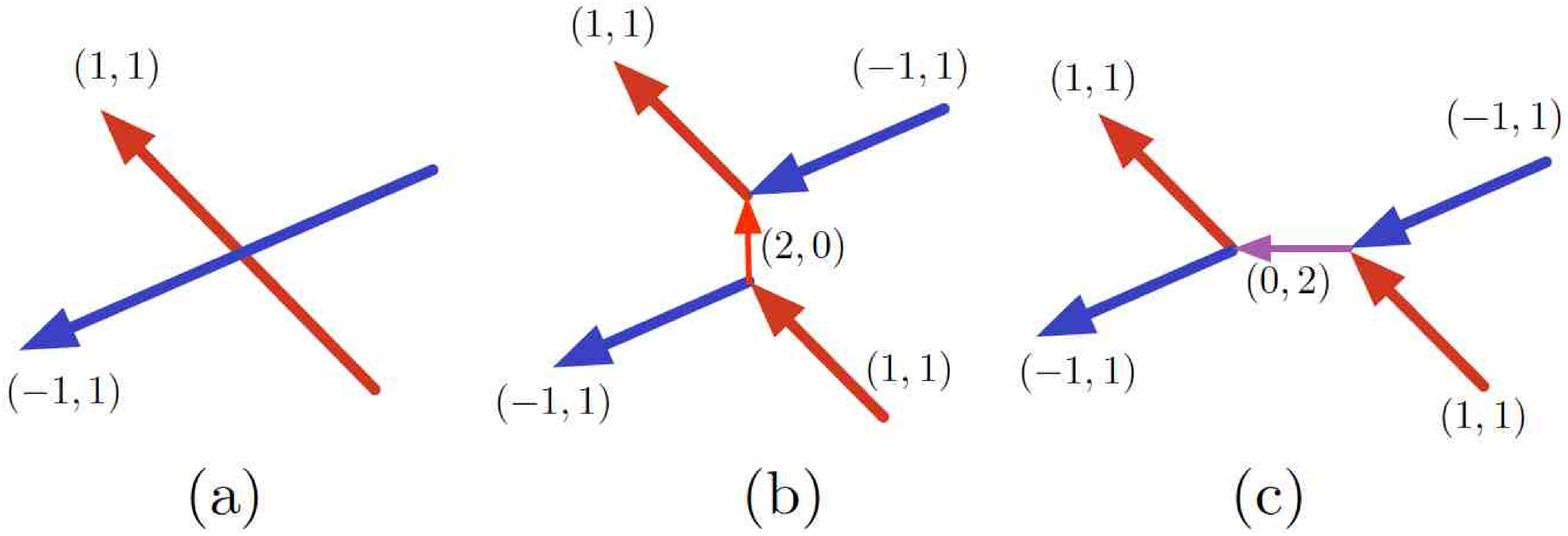,width=12cm} 
\caption{\small Relation between super D-Jackstraws and the string network. (a) The intersection point of two $(p,\,q)$-strings can be `regularized' into two $(p,\,q)$-junctions. There are two possible ways of the `regularization'. (b) The fundamental string mediates two string junctions. (c) The intermediate D-strings (with $(0,2)$ charge) are supposed to be along the direction of the negative $x^{1}$-axis to keep the supersymmetry.}
\label{figure10}}

In type IIB theory, there is another well-known configuration of $(p,\,q)$-strings; that is, the string network \cite{Sen:1997xi}\cite{Krogh:1997dx}. Since this configuration also preserves $8$ supersymmetries, one could conceive of its relation with the super D-Jackstraws. Each intersection point in the super $(p,\,q)$-Jackstraws can be `regularized' into two $(p,\,q)$-junctions. See Fig. \ref{figure10} for the schematic view.

%

\bigskip

\acknowledgments
The author thanks Soonkeon Nam for valuable comments and especially for adequately naming various bizarre configurations of D-branes.
This work was supported by the Science Research Center Program of the Korea Science and Engineering Foundation through 
the Center for Quantum Spacetime(CQUeST) of 
Sogang University with grant number R11 - 2005 - 021.
\begin{appendix}

\section{Static Intersecting D-strings}\label{appi}
\subsection{Supercharges in the Presence of D-branes}
The purpose of this section is to explain our tool used for the supersymmetry analysis throughout this paper. The supersymmetry preserved by a single Dp-brane is given by the sum of the left- and right-movers on the string world sheet, $Q_{\alpha}+(\beta_p^\bot \tilde{Q})_\alpha$  \cite{polchinski}. Here 
\begin{eqnarray}
\beta_p^\bot=\prod\limits_{m=p+1}^9 \beta^m,
\end{eqnarray}
and $\beta^m=\Gamma^m\Gamma$ is the spacetime parity operator on the world sheet of the open string. The chiral operator was denoted as $\Gamma=\prod^{9}_{i=0}\Gamma^{i}$.

As an application, let us consider a configuration of two static intersecting D-strings, one of which is along $x^{1}$-axis and the other is tilted at an angle $\theta$ with respect to the former in $(x^{1},\,x^{2})$-plane. The supercharges preserved by each D-string will be
\begin{eqnarray}\label{}
&&Q_{\alpha}+(\beta^{2}\beta^{\bot}_{2} \tilde{Q})_\alpha \nonumber\\
&&Q_{\alpha}+(\beta'^{2}\beta^{\bot}_{2} \tilde{Q})_\alpha,\qquad \beta'^{2}=\rho(-\theta)\beta^{2}\rho(\theta),
\end{eqnarray}
where we used the spinor representation of the rotation transformation as $\rho(\theta)=\exp{(\theta\Gamma^{12}/2)}$. 

The number of supersymmetries preserved by these two D-strings is the number of the spinor states invariant under the above two supercharge operators, that is, the number of $\mathbf{16}$-component chiral spinors, $\epsilon$, satisfying 
\begin{equation}\label{eq33}
\beta^{-2}\beta'^{2}\epsilon=\epsilon.
\end{equation}  

\subsection{$\mathbf{S}$-Representation for the Spinors}
To be explicit about solving the above spinor equation, we adopt $\mathbf{s}$-representation of the spinors used in Ref. \cite{polchinski}.
Spinors, being in the representation space of Clifford algebra, can be classified according to the eigenvalues of the following mutually commuting operators;
\begin{equation}\label{spinop}
2S_{0}=\Gamma^{0}\Gamma^{9},\quad 2S_{j}=-i\Gamma^{2j-1}\Gamma^{2j}\quad (j=1,\cdots 4).
\end{equation}    
Since all the operators are squared to $1$, they $(2S_{0},\,2S_{j})$ have two eigenvalues, $\pm1$. One can construct the raising or the lowering operators for each eigen value;
\begin{equation}\label{}
\Gamma^{0\pm}= \frac{1}{2} \left( \Gamma^{0}\pm\Gamma^{9} \right), \qquad \Gamma^{j\pm}= \frac{1}{2} \left(\Gamma^{2j-1}\pm i\Gamma^{2j} \right).
\end{equation}   
They indeed satisfy
\begin{equation}\label{}
[S_{0},\,\Gamma^{0\pm}]=\pm\Gamma^{0\pm},\quad [S_{j},\,\Gamma^{j\pm}]=\pm\Gamma^{j\pm}.
\end{equation}
Starting from the `ground state' $|0>$ defined by the condition that
\begin{equation}\label{}
\Gamma^{0-}|0>=\Gamma^{j-}|0>=0,
\end{equation}    
one can construct $32$-basis of the spinors by acting the raising operators $\Gamma^{0+}$ and $\Gamma^{j+}$ on the state $|0>$. According to their eigenvalues $2s_{0}$ and $2s_{j}$ for the operators $2S_{0}$ and $2S_{j}$, we represent the basis as
\begin{equation}\label{}
|2s_{0},\,2s_{1},\,2s_{2},\,2s_{3},\,2s_{4}>.
\end{equation}  
One thing to note is that it does not represent a `tensor' constructed via direct product, although all the operators (\ref{spinop}) mutually commute. Each entry of the basis is in one representation space of ten dimensional Clifford algebra. As a result, we have to keep track of the overall sign carefully in the computation.
For example, we will use the following results quite often:
\begin{eqnarray}\label{}
&&\Gamma^{0+}|-1,\,\pm1,\,2s_{2},\,2s_{3},\,2s_{4}>=|1,\,\pm1,\,2s_{2},\,2s_{3},\,2s_{4}>,\nonumber\\
&&\Gamma^{1+}|\pm1,\,-1,\,2s_{2},\,2s_{3},\,2s_{4}>=|\pm1,\,1,\,2s_{2},\,2s_{3},\,2s_{4}>.
\end{eqnarray}  

\subsection{Solving the Eigen-Spinor Equation}

Now we are ready to solve the eigenvalue equation (\ref{eq33}). Let
\begin{eqnarray}\label{epsilon}
\epsilon&=&a|2s_{0},\,1,\,2s_{2},\,2s_{3},\,2s_{4}>+b|2s_{0},\,-1,\,2s_{2},\,2s_{3},\,2s_{4}>\nonumber\\
&\equiv&(a,\,b),
\end{eqnarray}   
in terms of which the eigenvalue equation can be rewritten as  
\begin{equation}\label{finaleigen}
\left(\cos{\theta}-1 \right) \left( a,\,b\right) -i\sin{\theta} \left( a,\,-b\right)=0. 
\end{equation}  
In each term of the expression (\ref{epsilon}), the entries satisfy the chirality condition, say $\prod^{4}_{i=0}(2s_{i})=1$. The coefficients $a,\,b$ can be complex, which does {\it not} double the number of Killing spinors because $|2s_{0},\,2s_{1},\, \,2s_{2},\,2s_{3},\,2s_{4} >$ and $i|2s_{0},\,2s_{1},\, \,2s_{2},\,2s_{3},\,2s_{4} >$ are {\it not} independent.

Unless $\theta=0$, Eq. (\ref{finaleigen}) has no solution for $a$ and $b$. When $\theta=0$, that is when two D-strings are parallel, the equation is satisfied for arbitrary values of $a$ and $b$, thus results in 16 supersymmetries.

\end{appendix}

\end{document}